\documentclass[twocolumn,aps,superscriptaddress]{revtex4}
\usepackage{amssymb}
\usepackage{amsmath,bm}
\usepackage{graphicx}
\usepackage[normalem]{ulem}

\usepackage[usenames, dvipsnames]{xcolor}
\usepackage{lipsum}

\renewcommand{\sout}{\bgroup \color{red} \ULdepth=-.5ex \ULset}

\usepackage{isotope}

\begin{document}


\title{Bayesian inference on the isospin splitting of nucleon effective mass from giant resonances in $^{208}$Pb}

\author{Zhen Zhang\footnote{zhangzh275$@$mail.sysu.edu.cn}}
\affiliation{Sino-French Institute of Nuclear Engineering and Technology,
Sun Yat-sen University, Zhuhai 519082, China}

\author{Xue-Bin Feng}
\affiliation{Sino-French Institute of Nuclear Engineering and Technology,
Sun Yat-sen University, Zhuhai 519082, China}

\author{Lie-Wen Chen\footnote{lwchen$@$sjtu.edu.cn}}
\affiliation{School of Physics and Astronomy, Shanghai Key Laboratory for
Particle Physics and Cosmology, and Key Laboratory for Particle Astrophysics and Cosmology (MOE),
Shanghai Jiao Tong University, Shanghai 200240, China }

\date{\today}

\begin{abstract}
From a Bayesian analysis of the electric  dipole polarizability, the constrained energy of isovector giant dipole resonance, the peak energy of isocalar giant quadrupole resonance and the constrained energy of isocalar giant monopole resonance in $^{208}$Pb, we extract the isoscalar and isovector effective masses in nuclear matter at saturation density $\rho_0$ as 
$m_{s,0}^{\ast}/m=0.87_{-0.04}^{+0.04}$ and $m_{v,0}^{\ast}/m = 0.78_{-0.05}^{+0.06}$ at $90\%$ confidence level. The obtained constraints 
on $m_{s,0}^{\ast}$ and $m_{v,0}^{\ast}$ lead to  a positive isospin splitting of nucleon effective mass in asymmetric nuclear matter of isospin asymmetry $\delta$ at $\rho_0$ as $m_{n-p}^* / m=(0.20^{+0.15}_{-0.14})\delta$. In addition, the symmetry energy at the subsaturation density $\rho^{\ast}=0.05~\mathrm{fm}^{-3}$  is determined to be $E_{\mathrm{sym}}(\rho^{\ast})=16.7\pm1.3$ MeV at $90\%$ confidence level.
\end{abstract}

\maketitle


\section{\label{sec:level1}Introduction}
The nucleon effective mass, which characterizes the momentum or energy dependence of single nucleon potential in nuclear medium, is crucial in nuclear physics and astrophysics~\cite{JEUKENNE197683,MAHAUX19851,Jaminon1989,Li:2015pma,Li2018}. While various kinds of nucleon effective masses have been defined in nonrelativistic and relativistic approaches~\cite{MAHAUX19851,Jaminon1989,Li:2015pma, Li2018,Chen2007, Li2008}, here we focus on the total effective mass normally used in the nonrelativistic approach.
In asymmetric nuclear matter, effective masses of neutrons and protons, $m_n^{\ast}$ and $m_p^{\ast}$, may be different
due to the momentum dependence of the symmetry potential. The difference $m_{n-p}^{\ast}\equiv m_{n}^{\ast}-m_p^{\ast}$ is the so-called isospin splitting of nucleon effective mass, which plays an important role in many physical phenomena and questions in nuclear physics, astrophysics, and cosmology~\cite{Li:2015pma,Li2018}.
 For example, the $m_{n-p}^{\ast}$ affects the isospin dynamics in heavy-ion collisions~
 \cite{Li2004,Li2005,Chen:2004si, Rizzo2005,Giordano2010,Feng2013,
YXZhang2014,Xie2014},
thermodynamic properties of asymmetric nuclear matter~\cite{Behera_2011,Xu2015,Xu2015a}, and 
cooling of neutron stars~\cite{Baldo2014}.

Due to the limited isospin asymmetry in normal nuclei, the 
accurate determination of $m_{n-p}^{\ast}$ is rather difficult. Even the sign of $m_{n-p}^{\ast}$ remains a debated issue. For example, $m_{n-p}^{\ast}> 0$ in neutron-rich matter at the nuclear saturation density $\rho_0 \approx 0.16$~fm$^{-3}$ is favored by optical model analyses of nucleon-nucleus scattering data~\cite{LI2015408,Xu2010}, Skyrme energy density functional  (EDF)~\cite{Zhang2016a} and transport model~\cite{Kong2017} analyses of nuclear giant 
resonances, Brueckner-Hartree-Fock calculations~\cite{Zuo1999, Zuo2005, MA2004,Dalen2005}, chiral effective theory~\cite{Zhang2018c,Holt2013, Holt2016} and an analysis of various constraints on the magnitude and density slope of symmetry energy~\cite{Li2013a}, whereas the transport model analyses on single and/or double n/p ratio in heavy-ion collisions ~\cite{Coupland2016, Morfouace2019} (but see Ref.~\cite{Kong2015}) and an energy density functional study on nuclear electric dipole polarizability~\cite{Tuhin2018} lead to opposite 
conclusions. 

To understand these contradictive results and eventually determine the isospin splitting of nucleon effective mass needs  not only the improvement of both  theoretical models/calculations and experimental measurements, but also more sophisticated analysis approaches to 
quantifying the model uncertainties based on given experimental measurements. The latter 
is a quite general issue in nuclear theory, where due to the lack of a well-settled
\textit{ab initio} starting point, a lot of effective theories or models have been developed with parameters determined by fitting empirical knowledge or experimental data~\cite{Dobaczewski2014}. Over the past decade, various statistical approaches, e.g., covariance analysis~\cite{Reinhard2010,Piekarewicz2015}, Bayesian analysis~\cite{McDonnell2015,Bernhard2015, Pratt2015,Xie:2019sqb,Xie:2020tdo, Drischler:2020hwi,Xu2020,Kejzlar_2020} and bootstrap method~\cite{Bertsch2017,Pastore2019}, have been 
introduced in nuclear physics studies to quantify uncertainties and evaluate correlations of model parameters. Among them, the Bayesian inference method has been accepted as 
a powerful statistical approach and extensively used in various areas of nuclear physics. For a very recent review on Bayesian analysis and its applications in nuclear structure study, we refer the readers to Ref.~\cite{Kejzlar_2020}.

In our previous work~\cite{Zhang2016a}, we have extracted the isospin splitting of nucleon effective mass from the isovector giant dipole resonance (IVGDR) and isocalar giant quadrupole resonance (ISGQR) of $^{208}$Pb based on random phase approximation (RPA) calculations using a number of representative Skyrme interactions. However, some factors in the analysis, e.g., the choice of Skyrme interactions and the  priorly assumed linear  $1/E_{\mathrm{GQR}}^2$-$m_{s,0}^{\ast}$  relations, could affect the 
conclusions, and the statistical meaning of the obtained uncertainties are therefore unclear.
In the present work, within the framework of Skyrme energy density functional theory and 
random phase approximation approach, we employ the Bayesian inference method to 
extract the isospin splitting of nucleon effective mass from
the  electric dipole polarizability~\cite{Tamii2011a,Roca-Maza2015}, the constrained energy in IVGDR~\cite{Dietrich1988} and the ISGQR peak energy~\cite{Roca-Maza2013a} in $^{208}$Pb. The binding energy~\cite{Wang2017}, the charge radius~\cite{Angeli2013}, 
the constrained energy of isocalar giant monopole resonance (ISGMR)~\cite{Patel2013} and the neutron $3p_{1/2}-3p_{3/2}$ energy splitting~\cite{Vautherin1972} of $^{208}$Pb are also included in the analysis to guarantee the energy density functional can always reasonably describe the ground state and collective excitation state of $^{208}$Pb.  The isoscalar and isovector effective masses, and the neutron-proton effective mass splitting  at saturation density  together with the symmetry energy at the subsaturation density $\rho^{\ast}=0.05~\mathrm{fm}^{-3}$ are extracted from the Bayesian analysis.
 
The paper is organized as follows. In Sec.~\ref{Sec:Method}, we introduce the theoretical models and statistical approaches used  in this work. In the next section presented are the results for  the uncertainties of model parameters, the isospin splitting of nucleon effective mass at $\rho_0$ and the symmetry energy at $\rho^{\ast}=0.05~\mathrm{fm}^{-3}$. Finally, we draw conclusions in Sec.~\ref{Sec:Conclusions}.
\section{\label{Sec:Method}Model and Method}

\subsection{Nucleon effective mass in Skyrme energy density functional\label{Sec:SHF}}
As in Ref.~\cite{Zhang2016a}, we study the nucleon effective 
mass within the standard Skyrme energy density functional based on the conventional Skyrme interaction:
\begin{eqnarray}
v(\bm{r}_1,\bm{r}_2)&=&t_0(1+x_0P_{\sigma})\delta(\bm{r}_1-\bm{r}_2)   \notag \\
& &+\frac{1}{2}t_1(1+x_1P_{\sigma})[\bm{k}'^2\delta(\bm{r}_1-\bm{r}_2)+\mathrm{c.c.}] \notag \\
& &+t_2(1+x_2P_{\sigma})\bm{k}'\cdot\delta(\bm{r}_1-\bm{r}_2)\bm{k} \notag \\
& &+\frac{1}{6}t_3(1+x_3P_{\sigma})\rho ^{\alpha}\left(\frac{\bm{r}_1+\bm{r}_2}{2}\right)\delta(\bm{r}_1-\bm{r}_2)\notag\\
& &+iW_0(\bm{\sigma}_1+\bm{\sigma}_2)\cdot[\bm{k}'\times\delta(\bm{r}_1-\bm{r}_2)\bm{k}].
\label{Eq:Sky}
\end{eqnarray}
Here $\bm{\sigma}_i$ is the Pauli spin operator, $P_{\sigma}=(1+\bm{\sigma}_1\cdot\bm{\sigma}_2)/2$ is the spin-exchange operator, $\bm{k}=-i(\bm{ \nabla}_1-\bm{\nabla}_2)/2$ is the relative momentum operator, and $\bm{k}^{\prime}$ is the conjugate operator of $\bm{k}$ acting on the left.

Within the framework of Skyrme energy density functional, the  9  parameters $t_0- t_3$, $x_0- x_3$ and $\alpha$ of the Skyrme interaction can be expressed in terms of 9 macroscopic quantities
(pseudo-observables): the nuclear saturation density $\rho_0$, the energy per particle of symmetric nuclear matter $E_0(\rho_0)$, the incompressibility $K_0$, the isocalar effective mass $m_{s,0}^*$ at $\rho_0$, the isovector effective mass $m_{v,0}^*$ at $\rho_0$, the gradient coefficient $G_S$, the symmetry-gradient coefficient $G_{V}$, and the magnitude $E_{\mathrm{sym}}(\rho_0)$ and density slope $L$ of the nuclear symmetry energy at $\rho_0$~\cite{Chen2010, Chen:2011ps, Kortelainen2010}. The detailed analytical expressions can be found 
in Ref. \cite{Chen2010,Chen:2011ps}. Since the 9 macroscopic quantities have clear physical meaning and available empirical ranges, in the Bayesian analysis, we use them 
as model parameters. Consequently, our model has the following 10 parameters:
\begin{eqnarray}
\bm{p} & =& \lbrace  \rho_0,~E_0(\rho_0),~K_0,~E_{\mathrm{sym}}(\rho_0),~L, \notag \\
& &~ G_S,~G_V,~W_0,~m_{s,0}^{\ast},~m_{v,0}^{\ast}\rbrace.
\end{eqnarray}

In terms of the Skyrme parameter $t_0\sim t_3$ and $x_0\sim x_3$, the nucleon effective
mass in asymmetric nuclear matter with density $\rho $ and
isospin asymmetry $\delta$ can be expressed as~\cite{Chabanat1997}
\begin{eqnarray}
\label{Eq:EM}
\frac{\hbar ^2}{2m_{q}^{\ast}(\rho,\delta)}&=&\frac{\hbar ^2}{2m}
+\frac{1}{4}t_1\left[\left(1+\frac{1}{2}x_1\right)\rho
-\left(\frac{1}{2}+x_1\right)\rho_q\right] \notag \\
&+&\frac{1}{4}t_2\left[\left(1+\frac{1}{2}x_2\right)\rho
+\left(\frac{1}{2}+x_2\right)\rho_q\right].
\end{eqnarray}
The well-known isocalar and isovector effective masses, $m_{s}^{\ast}$
and $m_v^{\ast}$, which are respectively defined as the proton (neutron) effective mass 
in symmetric nuclear matter and pure neutron (proton) matter, are then given by~\cite{Chabanat1997}
\begin{eqnarray}
\label{Eq:ISEM}
  &\frac{\hbar^2}{2m_{s}^{\ast}(\rho)}=\frac{\hbar^2}{2m}+\frac{3}{16}t_1\rho+\frac{1}{16}t_2(4x_2+5)\rho, \\
  & \frac{\hbar^2}{2m_{v}^*(\rho)}=\frac{\hbar^2}{2m}+\frac{1}{8}t_1(x_1+2)\rho+\frac{1}{8}t_2(x_2+2)\rho.
\end{eqnarray}
Once given $m_{s}^{\ast}$
and $m_v^{\ast}$, the isospin splitting of nucleon effective mass can be 
obtained as~\cite{Zhang2015}
\begin{eqnarray}
\label{Eq:DEMNP}
m^{\ast}_{n-p}(\rho,\delta)&\equiv &\frac{m_{n}^{\ast}-m_{p}^{\ast}}{m}=2\frac{m_{s}^{\ast}}{m}\sum_{n=1}^{\infty} \left(\frac{m_{s}^{\ast}-m_{v}^{\ast}}{m_{v}^{\ast}}\delta\right)^{2n-1} \notag \\
&=& \sum_{n=1}^{\infty} \Delta m^*_{2n-1}(\rho)\delta^{2n-1},
\end{eqnarray}
with the isospin splitting coefficients $\Delta m^*_{2n-1}(\rho)$  given by
\begin{eqnarray}
\label{Eq:DmIso}
\Delta m^*_{2n-1}(\rho) = 2\frac{m_{s}^{\ast}}{m} \left(\frac{m_{s}^{\ast}}{m_{v}^{\ast}}-1\right)^{2n-1}.
\end{eqnarray}
In the following, we use $\Delta m_1^{\ast}$ to indicate the linear isospin splitting coefficient at the saturation density $\rho_0$.

\subsection{\label{Chap:GR} Nuclear giant resonances}

Nuclear giant resonances are usually studied  using random phase 
approximation (RPA) approach~\cite{Colo2013}.  For a given excitation operator $\hat{F}_{JM}$, the strength
function is calculated as:
\begin{equation}
S(E)=\sum_{\nu}|\langle\nu\Vert\hat{F}_J\Vert\tilde{0}\rangle|^2\delta(E-E_{\nu}),
\end{equation}
with $E_{\nu}$ being the energy of RPA excitation state $|\nu\rangle$,
and the 
moments $m_k$ of strength function (sum rules) are usually evaluated as:
\begin{equation}
m_k=\int dE E^kS(E)=\sum_{\nu}|\langle\nu\Vert\hat{F}_J\Vert\tilde{0}\rangle|^2E_{\nu}^k.
\end{equation}
For the ISGMR, IVGDR and ISGQR studied here, the excitation
operators are defined as:
\begin{eqnarray}
\hat{F}^{\mathrm{IS}}_0  & = & \sum_i^A r_i^2 \\
\hat{F}^{\mathrm{IV}}_{1M} &=& \frac{N}{A}\sum^Z_{i=1}r_iY_{\text{1M}}(\hat{r}_i)-\frac{Z}{A}\sum^N_{i=1}r_iY_{\text{1M}}(\hat{r}_i), \\
\hat{F}^{\mathrm{IS}}_{2M} &=& \sum_{i=1}^{A}r_i^2Y_{2M}(\hat{r}_i),
\end{eqnarray}
where $Z$, $N$ and $A$ are proton, neutron and mass number, respectively;
$r_i$ is the nucleon's radial coordinate; $Y_{\text{1M}}(\hat{r_i})$ and  $Y_{\text{2M}}(\hat{r_i})$ are the corresponding
spherical harmonic function.

Particularly, in linear response theory, the inverse energy weighted sum rule  can also be extracted from the constrained Hartree-Fock (CHF) approach~\cite{Bohigas1979,Sil2006}:
\begin{equation}
m_{-1} = -\left. \frac{1}{2}\frac{d^2\langle \lambda \vert \mathcal{H} \vert \lambda \rangle}{d\lambda^2}\right\vert_{\lambda=0},
\end{equation}
where $\vert\lambda\rangle$ is the ground-state for the nuclear system 
Hamilton $\mathcal{H}$ constrained by the field $\lambda \hat{F}_J$.

The energy of isoscalar giant monopole resonance (GMR), i.e., the breathing mode, is an important probe of the incompressibility in nuclear matter. It can be evaluated in the constrained approximation~\cite{BLAIZOT1980171} as 
\begin{equation}
E_{\mathrm{GMR}} =\sqrt{\frac{m_1(\mathrm{GMR})}{m_{-1}(\mathrm{GMR})}},
\end{equation}
where the energy weighted sum rule $m_1$ of ISGMR is related to the ground-state rms radius $\langle r^2 \rangle$ by~\cite{Bohr1975} 
\begin{equation}
m_1(\mathrm{GMR}) = 2 \frac{\hbar^2}{m}A\langle r^2 \rangle.
\end{equation}
Therefore, we calculate the $E_{\mathrm{GMR}}$ by using the CHF method for computational efficiency. 
 
For the isovector giant dipole resonance, we consider two observables, the electric dipole polarizability $\alpha_{\mathrm{D}}$ and the constrained energy $E_{\mathrm{GDR}}\equiv \sqrt{m_1/m_{-1}}$.  
The $\alpha_{\mathrm{D}}$ in $^{208}$Pb probes the symmetry energy at about $\rho_0/3$~\cite{Zhang2015} and is therefore sensitive to both the magnitude and  density slope of the symmetry energy at saturation density~\cite{Roca-Maza2013}. It is related to the inverse energy-weighted sum rule in the IVGDR by 
\begin{equation}
\alpha_{\mathrm{D}} = \frac{8\pi}{9}e^2m_{-1}(\mathrm{GDR}).
\end{equation}
Meanwhile, the energy weighted sum rule $m_1$ of IVGDR is related to 
the isovector effective mass at saturation density $m_{v,0}^{\ast}$ via~\cite{Colo2013,Zhang2016a}
\begin{equation}
m_{1}(\mathrm{GDR})=\frac{9}{4 \pi} \frac{\hbar^{2}}{2 m} \frac{N Z}{A}(1+\kappa)
\approx 
\frac{9\hbar^2}{8\pi}  \frac{N Z}{A} \frac{1}{m_{v,0}^{\ast}},
\end{equation}
where $\kappa$ is the well-known Thomas-Reiche-Kuhn sum rule enhancement. One then has 
approximately
\begin{equation}~\label{Eq:GDR}
E_{\mathrm{GDR}}^2 \propto \frac{1}{m_{v,0}^{\ast}\alpha_{\mathrm{D}}},
\end{equation}
which suggests that the $E_{\mathrm{GDR}}$ is negatively correlated to the 
$m_{v,0}^{\ast}$.

As well known, the excitation energy of isocalar giant quadruple resonance is sensitive to the isoscalar effective mass at saturation density. For example in the harmonic oscillator model, the ISGQR energy is~\cite{Bohr1975, Roca-Maza2013a}
\begin{equation}\label{Eq:GQR}
E_{\mathrm{GQR}}=\sqrt{\frac{2 m}{m_{s, 0}^{*}}} \hbar \omega_{0}
\end{equation}
with $\hbar \omega_0$ being the frequency of the harmonic oscillator.
In the present work,  we determine  the  $E_{\mathrm{GQR}}$ as 
the peak energy of the response function obtained from RPA calculations.
{To obtain continuous response function, the discrete RPA results are smeared out with Lorentzian functions. The width of Lorentzian functions is taken to be $3$ MeV to roughly reproduce experimental width $\sim3$ MeV of ISGQR.}

\begin{table}[htb]
\caption{Prior ranges of the ten parameters used.}
\label{Tab:Prior}
\begin{tabular}{lcc}
\hline \hline
Quantity & lower limit &  upper limit\\
\hline 
$\rho_0~(\mathrm{fm}^{-3})$                       &       0.155     &       0.165    \\
$E_0~(\mathrm{MeV}) $                             &     -16.5       &       -15.5      \\
$K_0~(\mathrm{MeV})$                              &      210.0      &       250.0       \\
$E_{\mathrm{sym}}(\rho_0)~(\mathrm{MeV})$         &      29.0        &      35.0  \\
$L~(\mathrm{MeV})$                                &      20.0         &      120.0    \\
$G_S~(\mathrm{MeV}\cdot \mathrm{fm}^{5})$          &     110.0        &       170.0   \\
$G_V~(\mathrm{MeV}\cdot \mathrm{fm}^{5})$          &     -70.0        &      70.0  \\
$W_0~(\mathrm{MeV}\cdot \mathrm{fm}^{5})$          &     110.0     &       140.0 \\
$m_{s,0}^{\ast}/m$                                &       0.7      &         1.0  \\
$m_{v,0}^{\ast}/m$                                &       0.6      &       0.9 \\
\hline\hline
\end{tabular}
\end{table}

\begin{table}[htb]
\caption{Experimental values and uncertainties used for the binding energy $E_B$~\cite{Wang2017}, the charge radius $r_C$~\cite{Angeli2013}, the breathing mode energy $E_{\mathrm{GMR}}$~\cite{Patel2013}, the neutron $3p_{1/2}-3p_{3/2}$ energy level splitting $\epsilon_{ls}$~\cite{Vautherin1972}, 
the electric dipole polarizability $\alpha_{\mathrm{D}}$~\cite{Tamii2011a,Roca-Maza2015}, the IVGDR constrained energy~\cite{Dietrich1988} and the ISGQR peak energy~\cite{Roca-Maza2013a}
 in $^{208}$Pb.}
\label{Tab:Data}
\begin{tabular}{lcc}
\hline \hline
 & value & $\sigma$ \\
\hline
$E_B$ (MeV)  & -1363.43  & 0.5 \\
$r_C$ (fm)   & 5.5012    & 0.01 \\
$E_{\mathrm{GMR}}$ (MeV) & 13.5 & 0.1 \\
$\epsilon_{ls}$ (MeV) &  0.89 & 0.09 \\
$\alpha_{\mathrm{D}}~(\mathrm{fm}^3)$ & 19.6 & 0.6 \\
$E_{\mathrm{GDR}}$ (MeV) & 13.46 & 0.1 \\
$E_{\mathrm{GQR}}$ (MeV) & 10.9  & 0.1 \\
\hline\hline
\end{tabular}
\end{table}

\subsection{Bayesian analysis}
The Bayesian analysis method has been widely accepted as a powerful statistical approach to quantifying the uncertainties and evaluating the correlations of model parameters as well as making predictions with certain confidence level according to experimental
measurements and empirical knowledge~\cite{McDonnell2015,Bernhard2015, Pratt2015,Xie:2019sqb,Xie:2020tdo, Drischler:2020hwi,Xu2020,Kejzlar_2020}. 
In this work, we employ the MADAI package~\cite{MADAI} to do Bayesian analysis based on the  Gaussian process emulators. For details on the statistical approach, we refer the readers to, e.g., Ref.~\cite{Bernhard2015}.

According to Bayes' theorem, the posterior probability distribution of model parameter $\bm{p}$ (which we are seeking for) given  experimental measurements
 $\bm{\mathcal{O}}^{\mathrm{exp}}$ for a set of observables $\bm{\mathcal{O}}$ can be evaluated as
\begin{equation}\label{Eq:Bay}
P(\bm{p} |\mathcal{M},\bm{\mathcal{O}}^{\mathrm{exp}})=\frac{P(\mathcal{M},\bm{\mathcal{O}}^{\mathrm{exp}}\mid \bm{p}) P(\bm{p})}{\int P(\mathcal{M},\bm{\mathcal{O}}^{\mathrm{exp}}\mid \bm{p}) P(\bm{p}) d \bm{p}},
\end{equation}
where $\mathcal{M}$ is the given model, $P(\bm{p})$ is the  prior probability of model parameters $\bm{p}$ before being confronted with the experimental measurements $\bm{\mathcal{O}}^{\mathrm{exp}}$, and $P( \mathcal{M},\bm{\mathcal{O}}^{\mathrm{exp}}|\bm{p})$ denotes the likelihood or the conditional probability of observing $\bm{\mathcal{O}}^{\mathrm{exp}}$ given the model $\mathcal{M}$ predictions
at $\bm{p}$. {The posterior univariate distribution of a single model parameter $p_i$ is given by 
\begin{equation}
P(p_i |\mathcal{M},\bm{\mathcal{O}}^{\mathrm{exp}})
=\frac{\int P(\bm{p}|\mathcal{M},\bm{\mathcal{O}}^{\mathrm{exp}})
d\prod_{j\neq i}p_j}{\int P(\bm{p} |\mathcal{M},\bm{\mathcal{O}}^{\mathrm{exp}})
d\prod_{j}p_j},
\end{equation}
and the correlated bivariate distribution of two parameters 
$p_i$ and $p_j$ is given by 
\begin{equation}
P[(p_i,p_j) |\mathcal{M},\bm{\mathcal{O}}^{\mathrm{exp}}]
=\frac{\int P(\bm{p}|\mathcal{M},\bm{\mathcal{O}}^{\mathrm{exp}})
d\prod_{k\neq i,j}p_k}{\int P(\bm{p} |\mathcal{M},\bm{\mathcal{O}}^{\mathrm{exp}})
d\prod_{k}p_k}.
\end{equation}
From the univariate distribution, the mean value of $p_i$ can be calculated as 
\begin{equation}
\langle p_i\rangle = \int p_i P(p_i |\mathcal{M},\bm{\mathcal{O}}^{\mathrm{exp}})dp_i.
\end{equation}
The confidence interval of $p_i$ at a confidence level $1-\alpha$ is normally obtained as the interval between the $(50\alpha)^{\text{th}}$ and $(100-50\alpha)^{\text{th}}$ percentile of 
the posterior univariate distribution. Particularly, the median value $U_{p_i}$ of $p_i$ is defined as the $50^{\mathrm{th}}$ percentile, i.e.,
\begin{equation}
\int^{U_{p_i}}_{-\infty} P(p_i |\mathcal{M},\bm{\mathcal{O}}^{\mathrm{exp}})dp_i = 0.5.
\end{equation}
}

For the prior distribution, we assume the ten parameters uniformly distributed in the empirical ranges listed in Tab.~\ref{Tab:Prior}. {As can be seen from Eq.~(\ref{Eq:Bay}), the posterior 
distribution is determined by the combination of the prior distribution and 
likelihood function, which depends on the experimental measurement for 
observable. Therefore, the prior distribution is very important in the Bayesian analysis and can significantly affect the extracted constraints. Nevertheless, in the present work, due to the relatively poor knowledge on $m_{s,0}^{\ast}$ and $m_{v,0}^{\ast}$ in the Skyrme EDF, we assume large prior ranges for 
$m_{s,0}^{\ast}$ and $m_{v,0}^{\ast}$ and thus the constraints on nucleon effective masses are mainly due to the giant resonance 
observables. Narrowing the prior ranges of all parameters by $20\%$ only slightly reduces the posterior uncertainties of the iospin splitting of nucleon effective mass by few percent.}

The likelihood function is taken to be the commonly used Gaussian form 
\begin{equation}\label{Eq:Likelihood}
P(\mathcal{M},\bm{\mathcal{O}}^{\mathrm{exp}}_i\mid \bm{p}) \propto\mathrm{exp} \left\{
-\sum_i\frac{[\mathcal{O}_i(\bm{p})-\mathcal{O}^{\mathrm{exp}}_i]^2}{2\sigma_i^2}
 \right\},
 \end{equation}
 where $\mathcal{O}_i(\bm{p})$  is the model prediction for an observable at given point $\bm{p}$, $\mathcal{O}^{\mathrm{exp}}_i$ is the corresponding experimental measurement, and $\sigma_i$ is the uncertainty or the width of likelihood function. For a given parameter set  $\bm{p}$, we calculate 
the following 7 observables in $^{208}$Pb from Hartree-Fock, CHF and RPA calculations:
the electric dipole polarizability $\alpha_{\rm{D}}$~\cite{Tamii2011a,Roca-Maza2015}, the IVGDR constrained energy $E_{\rm{GDR}}$\cite{Dietrich1988}, the ISGQR peak energy $E_{\mathrm{GQR}}$~\cite{Roca-Maza2013a}, the binding energy $E_B$~\cite{Wang2017}, 
the charge radius $r_{C}$~\cite{Angeli2013}, the breathing mode energy $E_{\mathrm{GMR}}$~\cite{Patel2013} and the neutron $3p_{1/2}-3p_{3/2}$ energy level splitting $\epsilon_{ls}$~\cite{Vautherin1972}. The experimental values for the 7 observables together with the assigned 
uncertainties are listed in Tab.~\ref{Tab:Data}. For the $\alpha_{\mathrm{D}}$ and  $E_{\mathrm{GMR}}$, the  $\sigma_i$ are taken to be their 
experimental uncertainties given in Refs.~\cite{Tamii2011a} and~\cite{Patel2013}; for the well determined $E_B$, $r_C$, $\epsilon_{ls}$ and $E_{\mathrm{GDR}}$, we assign them artificial $1\sigma$ errors of $0.5$ MeV,  $0.01$ fm, $0.09$ MeV and $0.1$ MeV, respectively; for the experimental value and uncertainty of $E_{\mathrm{GQR}}$, we use the  weighted average of
experimental measurements, $10.9\pm0.1$ MeV reported in Ref.~\cite{Roca-Maza2013a}. We note that decreasing the artificial errors of $E_B$, $r_C$, $\epsilon_{ls}$ and $E_{\mathrm{GDR}}$ by half only slightly reduces the posterior uncertainty of  $\Delta m_1^*$ by about $7\%$, and does not affect the constraint on the symmetry energy at the subsaturation density $\rho^*=0.05~\mathrm{fm}^{-3}$.

According to the prior distribution and defined likelihood function, 
the Markov chain Monte Carlo (MCMC) process using Metropolis-Hastings algorithm is performed to 
evaluate the posterior distributions of model parameters. For 10-dimensional parameter space in this work, a huge number of MCMC steps are needed to extract posterior distributions, and thus theoretical calculations for all MCMC steps are infeasible. Instead, in this work, we 
first sample a number of parameter sets in the designed parameter space,
and train  Gaussian process (GP) emulators~\cite{GP} using the model predictions with the sampled parameter sets. The obtained GPs provide fast interpolators and are used to evaluate the likelihood function in each MCMC step.

\begin{figure}[htb]

\includegraphics[width=0.95\linewidth]{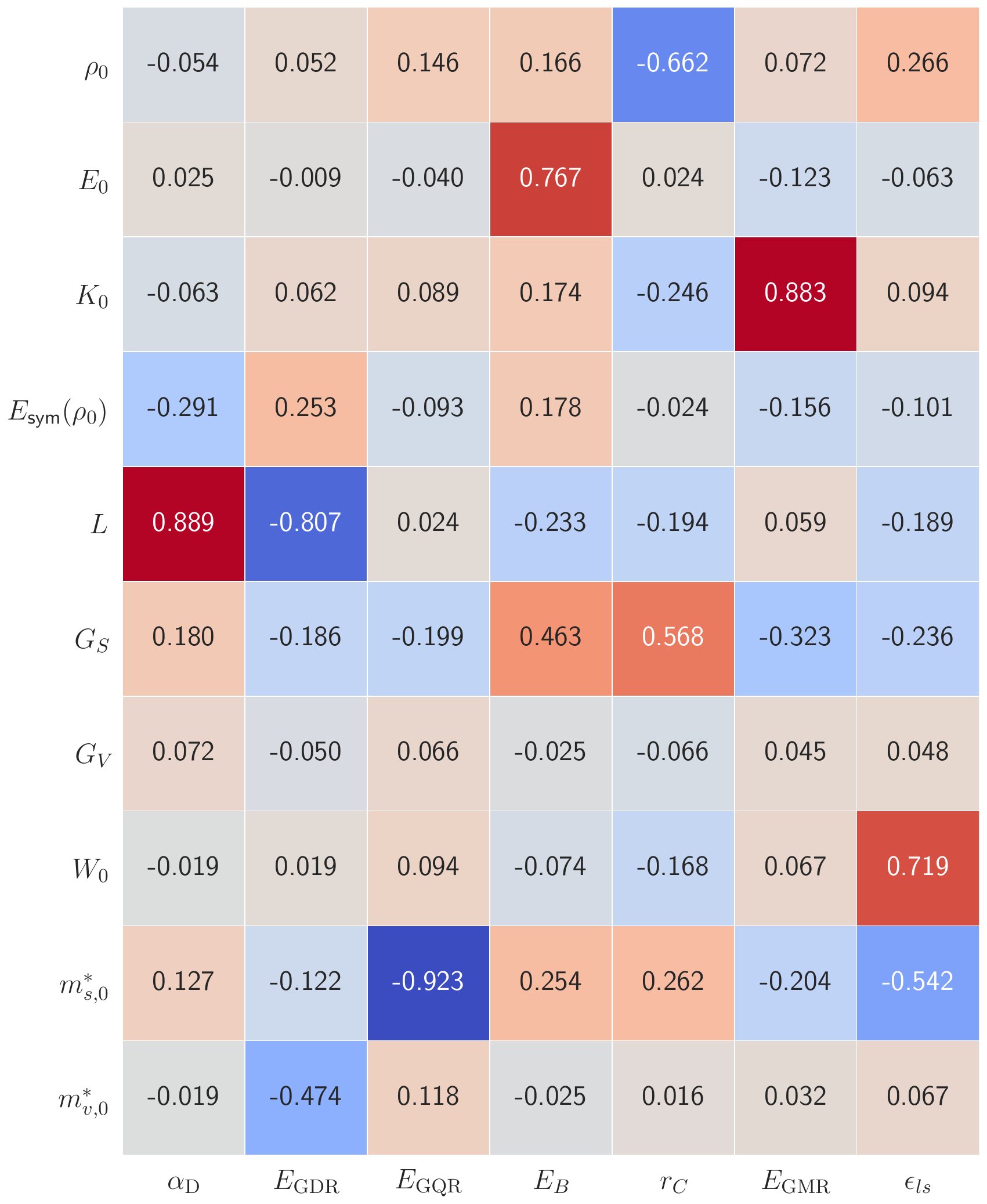}

\caption{(Color online). Visualization of the Pearson correlation 
coefficients  among model parameters and observables from the training data. The numerical values of correlation coefficients are annotated and color-coded, where darker red indicates more positive values and darker blue indicates more negative.}
\label{Fig:Correlation}
\end{figure}

\begin{figure*}[htb]
\centering
\includegraphics[width=\linewidth]{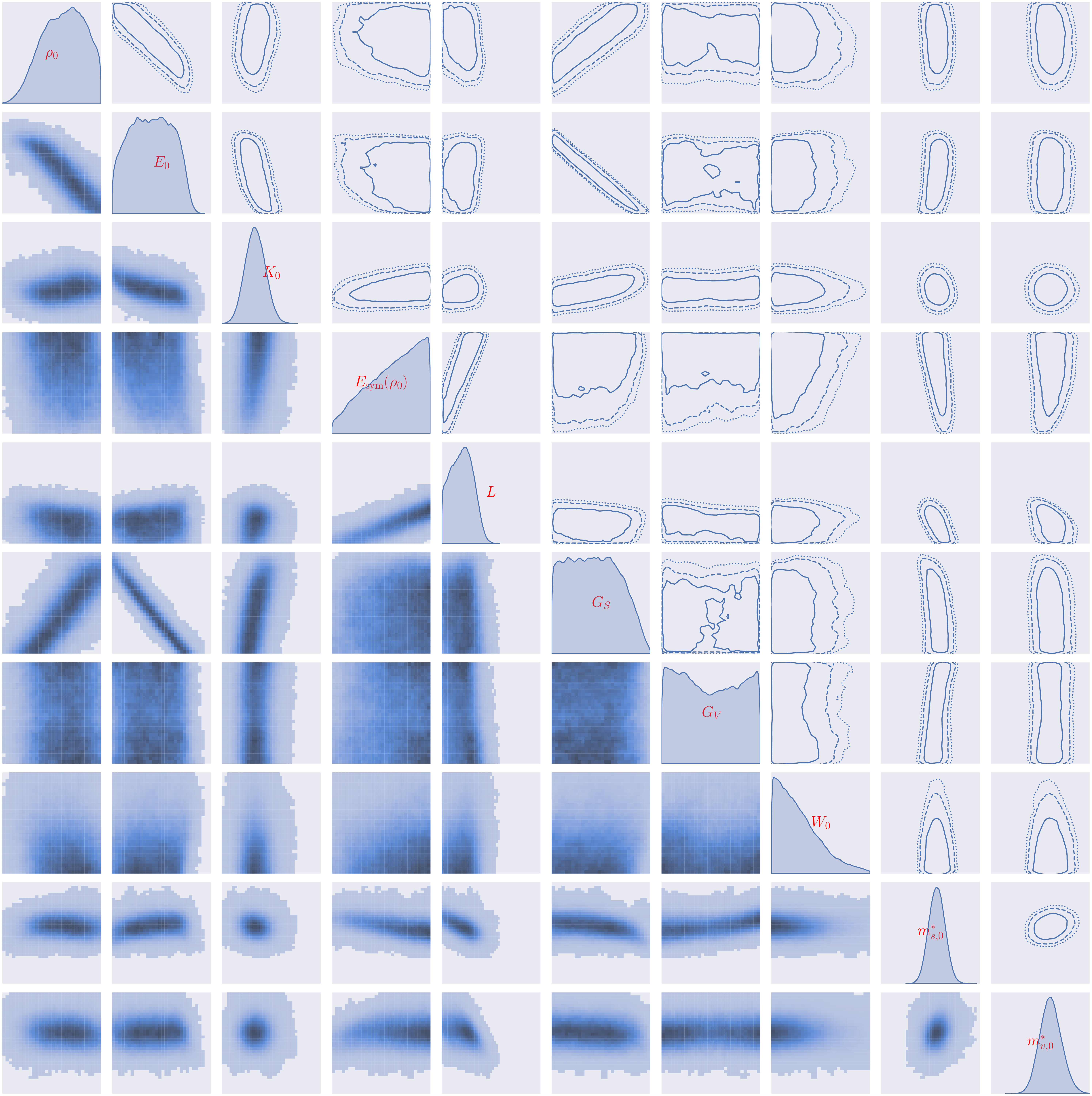}

\caption{(Color online). Univariate and bivariate  posterior distributions of the 10 model parameters. The solid, dashed and dotted lines in the upper-right panels enclose $68.3\%$, $90\%$ and $95.4\%$ confidence regions, respectively; the diagonal plots show Gaussian kernel density estimations of the posterior marginal distribution for respective parameters; in the lower-left panels are scatter histograms of MCMC samples.
The range of variations of the 10 parameters are the prior range listed in Table~\ref{Tab:Prior}.}
\label{Fig:Post}
\end{figure*}

\begin{table*}[htb]
\caption{Best value, mean, median and confidence intervals of the model parameters from MCMC sampling.}
\label{Tab:Post}
\begin{tabular}{lcccccc}
\hline \hline
& best & mean & median & $68.3\%$ C.I. &  $90\%$ C.I. & $95.4\%$ C.I. \\
\hline
$\rho_0~(\text{fm}^{-3})$& 0.1597 & 0.1612 & 0.1613 & $0.1589\sim0.1635$ & $0.1577\sim0.1644$ & $0.1570\sim0.1647$\\
$E_0~(\text{MeV})$& -16.04 & -16.10 & -16.10 & $-16.34\sim-15.87$ & $-16.44\sim-15.78$ & $-16.47\sim-15.74$ \\
$K_0~(\text{MeV})$& 224.6 & 223.5 & 223.4 & $219.4\sim227.6$ & $216.9\sim230.3$ & $215.6\sim231.8$ \\
$E_{\text{sym}}(\rho_0)~(\text{MeV})$& 34.4 & 32.7 & 33.0 & $30.9\sim34.4$ & $29.9\sim34.8$ & $29.5\sim34.9$ \\
$L(\rho_0)~(\text{MeV})$& 48.8 & 40.3 & 40.4 & $27.9\sim51.9$ & $22.8\sim58.1$ & $21.4\sim61.1$ \\
$G_S~(\text{MeV}\cdot \text{fm}^5)$& 125.7 & 135.5 & 135.1 & $118.2\sim152.5$ & $112.7\sim160.3$ & $111.2\sim163.5$ \\
$G_V~(\text{MeV}\cdot \text{fm}^5)$& 65.0 & -1.6 & -3.1 & $-50.9\sim49.5$ & $-64.1\sim63.9$ & $-67.3\sim67.3$ \\
$W_0~(\text{MeV}\cdot \text{fm}^5)$& 111.6 & 118.4 & 117.0 & $112.0\sim125.1$ & $110.6\sim131.4$ & $110.3\sim134.7$ \\
$m_{s,0}^*/m$& 0.88 & 0.87 & 0.87 & $0.84\sim0.89$ & $0.83\sim0.91$ & $0.82\sim0.92$ \\
$m_{v,0}^*/m$& 0.78 & 0.78 & 0.78 & $0.75\sim0.81$ & $0.73\sim0.84$ & $0.72\sim0.85$ \\ 
\hline\hline
\end{tabular}
\end{table*}

\section{\label{Sec:Result}Results and Discussions }
We first generated 2500 parameter sets using the maximin Latin cube sampling method~\cite{Morris1995}. 24 of them are near the edge of the allowed parameter space and lead to 
numerical instability in Hartree-Fock (HF) or 
RPA calculations. Therefore, we discarded the 24 parameter sets and used
the left 2476 parameter sets in HF, CHF and RPA calculation to obtain the 
training data for Gaussian emulators.

Results from the 2476 training points can inform us of correlations between observables and model parameters. 
We show in Fig.~\ref{Fig:Correlation}  the Pearson correlation coefficients among model parameters and the chosen 7 observables obtained from the training data. 
In Fig.~\ref{Fig:Correlation}, darker red indicates positive larger value and thus stronger positive correlation, and 
darker blue indicates more negative value and thus stronger negative correlation. One can expect that the parameters strongly correlated with the chosen observables are more likely to be constrained. Particularly, it is seen that the correlations among observables  in nuclear giant resonances
and model parameters are nicely consistent with the empirical knowledge introduced in Chapter~\ref{Chap:GR}: the $\alpha_{\mathrm{D}}$ is positively (negatively) correlated to  $L$ [$E_{\mathrm{sym}}(\rho_0)$] because it is mostly sensitive to the symmetry energy at $\rho^{\ast}=0.05~\mathrm{fm}^{-3}$~\cite{Zhang2015};  the $E_{\mathrm{GDR}}$ is negatively correlated to  both $m_{v,0}^{\ast}$
and $L$, but positively to $E_{\mathrm{sym}}(\rho_0)$, which can be understood from Eq.~(\ref{Eq:GDR}) and the dependence of $\alpha_{\mathrm{D}}$ on $L$ and $E_{\mathrm{sym}}(\rho_0)$; the $E_{\mathrm{GQR}}$ is strongly negatively correlated with $m_{s,0}^{\ast}$  [see Eq.~(\ref{Eq:GQR})]; the $E_{\mathrm{GMR}}$ is mostly sensitive to the incompressibility $K_0$. Meanwhile, the $G_V$ is weakly correlated to all observables, and therefore can not 
be well constrained in the present analysis.

Based on the training data, Gaussian process emulators were tuned to quickly 
predict model output for MCMC process. With the help of GPs, we first ran $10^6$ burn-in MCMC steps to allow the chain to reach equilibrium,  and then generated $10^7$ points in parameter space via the MCMC sampling. The posterior distributions of model parameters are  then extracted from the $10^7$ samples, and visualized in Fig.~\ref{Fig:Post}. The lower-left panels show the bivariate  scatter histograms 
of the MCMC samples; the diagonal ones present the univariate posterior 
distribution of the model parameters; the solid, dashed, and dotted lines in the upper-right panels enclose $68.3\%$,  $90\%$ and $95.4\%$ confidence regions, respectively. Fig.~\ref{Fig:Post} intuitively present the uncertainties and correlations of the model parameter imposed by the experimental measurements for the chosen observables. Remarkably, $K_0$, $m_{s,0}^{\ast}$, and $m_{v,0}^{\ast}$ are well constrained by the giant monopole, dipole and quadruple resonances, respectively. Another interesting feature is the strong positive correlation between $E_{\mathrm{sym}}(\rho_0)$ and $L$, 
which can be understood by the fact that $\alpha_{\mathrm{D}}$  is positively correlated to $L$ but negative to $E_{\mathrm{sym}}(\rho_0)$ (see Fig.~\ref{Fig:Correlation}). Similarly, the $E_B$ datum leads to a negative $E_0$-$G_S$ correlation, and $r_C$ datum results in 
a positive $\rho_0$-$G_S$ correlation. Therefore, the combination of $E_B$ and $r_C$ leads to a negative $\rho_0$-$E_0$ correlation. 
\begin{figure}[htb]
\centering
\includegraphics[width=\linewidth]{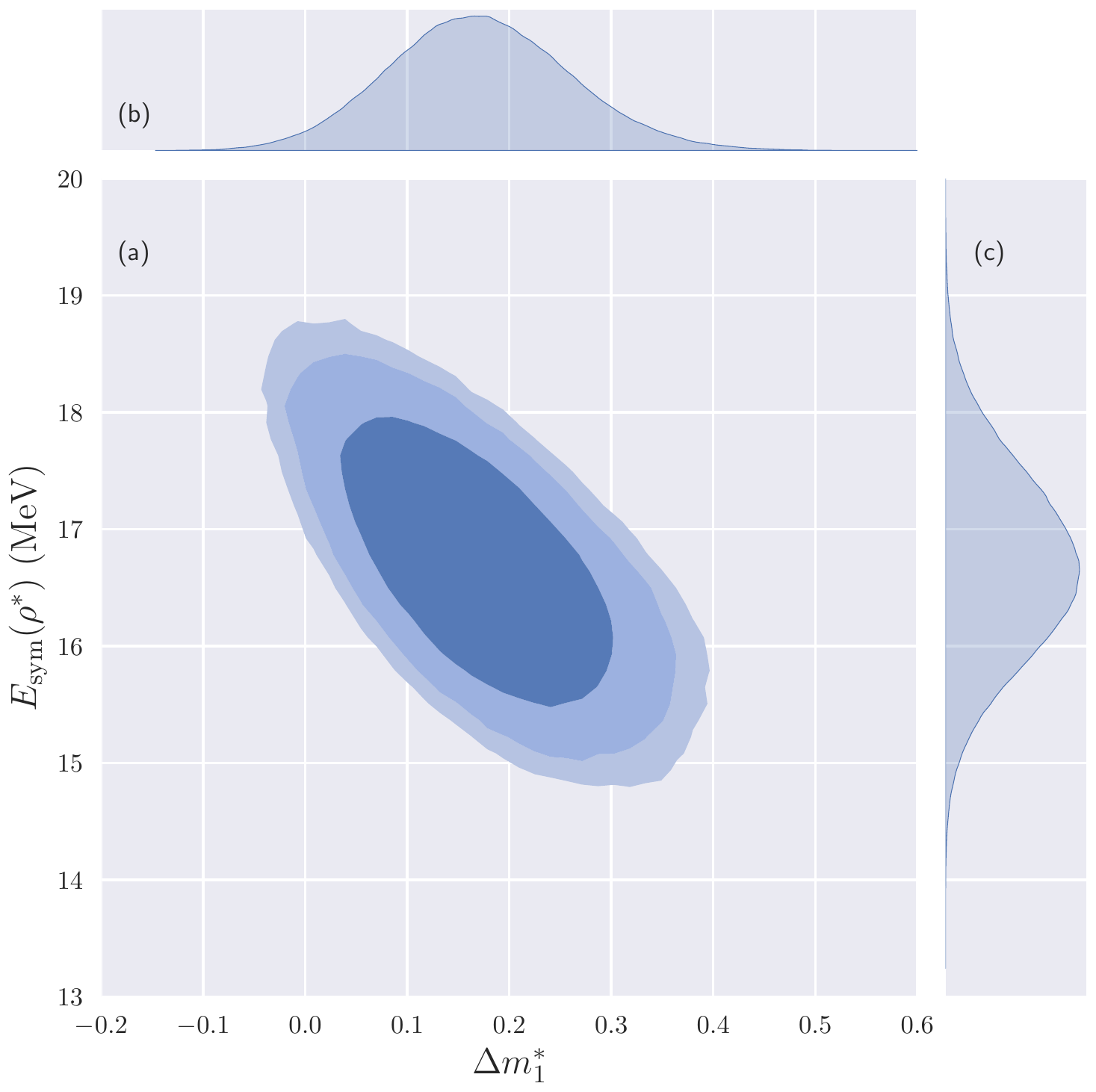}
\caption{(Color online). Posterior {bivariate} (a) and {univariate} [(b) and (c)]   distributions of $E_{\mathrm{sym}}(\rho^{\ast})$ at $\rho^{\ast}=0.05~\mathrm{fm}^{-3}$ and the linear isospin splitting coefficient $\Delta m_1^{\ast}$ at $\rho_0$. The shaded regions in window (a) indicate the $68.3\%$, $90\%$ and $95.4\%$ confidence regions. }
\label{Fig:S05Dm1}
\end{figure}

To quantify the posterior distribution from the Bayesian analysis, we 
list statistical quantities estimated from the MCMC samples, including 
the mean value, median value and confidence intervals at $68.3\%$, $90\%$ and $95.4\%$ confidence levels.  The best value, i.e., the parameter set that gives the largest likelihood function, is also listed for reference. 
In particular, we obtain $m_{s,0}^{\ast}/m=0.87_{-0.03}^{+0.02}$ and $m_{v,0}^{\ast}/m = 0.78_{-0.03}^{+0.03}$ at $68\%$ confidence level, and 
$m_{s,0}^{\ast}/m=0.87_{-0.04}^{+0.04}$ and $m_{v,0}^{\ast}/m = 0.78_{-0.05}^{+0.06}$ at $90\%$ confidence level. The results are consistent with $m_{s,0}^{\ast}=0.91\pm0.05$ and $m_{v,0}^{\ast}/m =0.8\pm0.03$ extracted from the GDR and GQR in Ref.~\cite{Zhang2016a} using the conventional method. {We would like to mention that 
that compared with our previous work in Ref.~\cite{Zhang2016a}, where a conventional analysis was carried out based on 
only 50 representative Skyrme EDFs, the present work extract the posterior distributions of model parameters from 
a huge number of parameter sets from MCMC sampling. The uncertainties of model parameters are thus better evaluated and the constraints obtained in the present work should be more reliable.}
The obtained $90\%$ confidence interval for $m_{v,0}^{\ast}$ is also in very good agreement with the result of $0.79_{-0.06}^{+0.06}$ from a recent 
Bayesian analysis of giant dipole resonance in $^{208}$Pb~\cite{Xu2020}. Note that
compared with the present work, the
Bayesian analysis in  Ref.~\cite{Xu2020} uses the same GDR data,
but the MCMC process is based on fully self-consistent RPA calculations. Therefore, the consistence between the two results also verifies the reliability of Gaussian emulators as a fast surrogate of  real model calculations.

In Fig.~\ref{Fig:S05Dm1}, we further show the posterior bivariate and univariate distributions of      the symmetry energy at $\rho^{\ast}=0.05~\mathrm{fm}^{-3}$ and the linear isospin splitting coefficient $\Delta m_1^{\ast}$ at $\rho_0$. 
Because of the approximate relations  $E_{\mathrm{sym}}(\rho^{\ast}) \propto 1/\alpha_{\mathrm{D}}$ and
$E_{\mathrm{GDR}}^2\propto (\alpha_{\mathrm{D}}m_{v,0}^{\ast})^{-1}$, the GDR data lead to positive correlation
 between $E_{\mathrm{sym}}(\rho^{\ast})$ and $m_{v,0}^{\ast}$. Therefore, Fig.~\ref{Fig:S05Dm1} exhibits a negative $E_{\mathrm{sym}}(\rho^{\ast})$-$\Delta m_1$ correlations [see Eq.~(\ref{Eq:DmIso})]. 
The confidence intervals of $E_{\mathrm{sym}}(\rho^{\ast})$ and $\Delta m^*_{1}$  can be extracted from their univariate distributions shown in 
Fig.~\ref{Fig:S05Dm1} (b) and (c). Specifically, we obtained $E_{\mathrm{sym}}(\rho^{\ast}) = 16.7_{-0.8}^{+0.8}~\mathrm{MeV}$ and
$\Delta m_1^{\ast} = 0.20_{-0.09}^{+0.09}$ at $68.3\%$ confidence level, and $E_{\mathrm{sym}}(\rho^{\ast}) = 16.7_{-1.3}^{+1.3}~\mathrm{MeV}$ and 
$\Delta m_1^{\ast} = 0.20_{-0.14}^{+0.15}$ at $90\%$ confidence level. For the higher order terms, we find, for example, $\Delta m_3^{\ast}$ is less than 0.01 at $90\%$ confidence level and therefore can be neglected.

Within the uncertainties, the present constraint on $\Delta m_1^{\ast}$ is consistent with the constraints $m_{n-p}^{\ast}/m = (0.32\pm0.15)\delta$ ~\cite{Xu2010} and $m_{n-p}^{\ast}/m  = (0.41\pm0.15)\delta$~\cite{LI2015408} 
extracted from  the global optical model analysis of nucleon-nucleus 
scattering data, and also agrees with $\Delta m_1^{\ast}=0.27$ obtained by analysing various constraints on the magnitude and density slope of the symmetry energy~\cite{Li2013a}. It is also consistent with the constraints from analyses of isovector GDR and isocalar GQR with RPA calculations using Skyrme interactions~\cite{Zhang2016a} and transport model using an improved isospin- and momentum-dependent interaction~\cite{Kong2017}. In addition, the present constraint $E_{\mathrm{sym}}(\rho^*) = 16.7^{+1.3}_{-1.3}$~MeV is consistent with the result $15.91 \pm 0.99$~MeV obtained in Ref.~\cite{Zhang2015} where the data on the $\alpha_{\mathrm{D}}$ in $^{208}$Pb does not consider the contribution of the quasideuteron effect~\cite{Roca-Maza2015}. Including the quasideuteron effect will slightly enhance the $E_{\mathrm{sym}}(\rho^*)$ and thus improves the agreement with the present result.

To end the section, we would like to discuss the limitations of the present work. In this work we only focus on nuclear giant resonances in $^{208}$Pb. However, the $m_{v,0}^{\ast}$ from the GDR of $^{208}$Pb is not consistent with the GDR in $^{16}$O~\cite{Erler_2010}. Describing the giant  resonances simultaneously  in light and heavy nuclei is still a challenge.
For the ambiguities in determining nucleon effective masses from nuclear giant resonances, we refer the readers to Ref.~\cite{Li2018}. It is also worth mentioning that due to the simple quadratic momentum dependence of the single-nucleon potential in Skyrme energy density functional, the nucleon effective mass is momentum independent and only has a simple density dependence[see Eq.~(\ref{Eq:EM})], which is not the case in microscopic many body theories like chiral effective theory~\cite{Holt2013, Holt2016}. The extended Skyrme pseudopotential~\cite{Carlsson2008,Raimondi2011,Wang2018} with higher order momentum-dependent terms may help to address the issues on the isospin splitting of nucleon effective mass.

\section{\label{Sec:Conclusions} Conclusions}

Within the framework of Skyrme energy density functional and random phase approximation, 
we have done a Bayesian analysis for
the data on the ground and collective excitation states of $^{208}$Pb to extract information on the nucleon effective mass and its isospin splitting. Our results indicate that the isoscalar effective mass $m^*_{s,0}/m$ exhibits a particularly strong correlation with the peak energy of isocalar giant quadrupole resonance, and the isovector effective mass $m^*_{v,0}/m$ is correlated with the constrained energy of isovector giant dipole resonance.
 By including in the analysis the constrained energy of the isoscalar monopole resonance, the peak energy of isocalar giant quadrupole resonance, the electric dipole polarizability
 and the constrained energy of the isovector giant dipole resonance, we have constrained the isocalar and isovector  effective masses,  and 
the isospin splitting of nucleon effective mass at saturation density, respectively, as
$m_{s,0}^{\ast}/m =0.87^{+0.04}_{-0.04} $, 
$m_{v,0}^{\ast}/m =0.78^{+0.06}_{-0.05} $, and $m_{n-p}^{\ast}/m = (0.20_{-0.14}^{+0.15})\delta$ at $90\%$ confidence level.
Corresponding to the $68.3\%$ ($1\sigma$) confidence level, the constraints become $m^*_{s,0}/m =0.87^{+0.02}_{-0.03}$, $m^*_{v,0}/m =0.78^{+0.03}_{-0.03}$, and $ m_{n-p}^{\ast}/m = (0.20_{-0.08}^{+0.09})\delta$. In addition, the symmetry energy at the subsaturation density $\rho^{\ast}=0.05~\mathrm{fm}^{-3}$ has been constrained  as  $E_{\mathrm{sym}}(\rho^{\ast}) = 16.7_{-0.8}^{+0.8}~\mathrm{MeV}$ at $68.3\%$ confidence level, and $E_{\mathrm{sym}}(\rho^{\ast}) = 16.7_{-1.3}^{+1.3}~\mathrm{MeV}$ at $90\%$ confidence level.

\section*{Acknowledgements}

This work was supported in part by the National Natural Science Foundation of China under Grants No. 11905302 and
No. 11625521, and   National SKA Program of China No. 2020SKA0120300.

\bibliography{ref.bib}
\end{document}